# KNOWLEDGE MANAGEMENT STRATEGIES AND PROCESSES IN AGILE SOFTWARE DEVELOPMENT: A SYSTEMATIC LITERATURE REVIEW

RAQUEL ANDRADE BARROS OURIQUES*,‡, KRZYSZTOF WNUK*,¶, TONY GORSCHEK*,§,
RICHARD BERNTSSON SVENSSON*,||,†,**

*Software Engineering Department, Blekinge Institute of Technology, Valhallavägen 1,
Karlskrona, 37142,Sweden
‡ raquel.ouriques@bth.se
¶ krzysztof.wnuk@bth.se
§ tony.gorschek@bth.se
|| richard.berntsson.svensson@bth.se

†Department of Computer Science and Engineering, Chalmers | University of Gothenburg
Gothenburg, Sweden
** richard.berntsson.svensson@gu.se

Knowledge-intensive companies that adopt Agile Software Development (ASD) relay on efficient implementation of Knowledge Management (KM) strategies to promotes different Knowledge Processes (KPs) to gain competitive advantage. This study aims to explore how companies that adopt ASD implement KM strategies utilizing practices that promote the KPs in the different organizational layers. Through a systematic literature review, we analyzed 32 primary studies, selected by automated search and snowballing in the extant literature. To analyze the data, we applied narrative synthesis. Most of the identified KM practices implement personalization strategies (81%), supported by codification (19%). Our review shows that the primary studies do not report KM practices in the strategic layer and two of them in the product portfolio layer; on the other hand, in the project layer, the studies report 33 practices that implement personalization strategy, and seven practices that implement codification. KM strategies in ASD promote mainly the knowledge transfer process with practices that stimulates social interaction to share tacit knowledge in the project layer. As a result of using informal communication, a significant amount of knowledge can be lost or not properly transferred to other individuals and, instead of propagating the knowledge, it remains inside a few individuals' minds.

## 1. Introduction

Software-intensive product development companies struggle to stay competitive due to fierce competition and increased pressure forces to quickly release new products to the market [1][2]. These forces push companies to improve resource management aiming for better product quality, creativity or efficient development process [3]. Many companies have also introduced Agile Software Development (ASD) methods, as they are perceived to offer better response to frequently changing market needs, more flexible software development methods and shorter learning cycles [4].

The increased flexibility that ASD promises comes with the cost of prioritizing informal communication between team members over written documentation [5] or loss of the "big picture" of the product due to extensive focus on developing features [6]. As a result, managing the knowledge asset becomes critical, and the lack of it can lead to several negative effects. For example, barriers to collaboration and asynchronous communication in large companies that introduced ASD [7], and competitiveness loss [8].

Mastering how to manage knowledge as a competitive asset [9] can improve team communication and the change responsiveness. The utilization of knowledge management (KM) strategies can significantly support achieving and sustaining competitive advantage [10][11] and brings several benefits to software development [12][13], e.g. that effective networks for tacit knowledge sharing in ASD contribute to continuous process improvement [14].

How to manage knowledge in ASD is still not well understood or investigated [15]. Rus and Lindvall [12] affirm that the reason for failures in KM deployments happens because many companies do not establish their goals in the KM strategy, and manage documents instead of knowledge.

KM focuses on the effort to manage an organizations' workforce through social processes that facilitate interaction between individuals [11]. KM also has established principles for dynamic activities that focus on the knowledge processes (KPs): creation, storage/retrieval, transfer, and application [9].

Knowledge processes often use practices that implement *codification* and/or *personalization* strategies; which should be rooted in the company's corporate strategy [16], for example, physical open spaces, which implement a personalization strategy, could promote knowledge transfer between individuals in physical structure that facilitates the working together [17]. Although the research literature in ASD encompass several studies on practices for specific KPs [16][17]; what is lacking, however, is understanding how KM strategies relate to the different KPs in companies adopting ASD.

While the software engineering field had addressed the storage and retrieval knowledge process, the field is still distant from the KM mainstream research, and processes such as knowledge creation, transfer and application remain largely unexplored [15]. This study contributes to filling this gap by means of focusing on how companies adopting ASD manage the knowledge asset implementing KM strategies, focusing on the KPs that they promote in the different organizational layers.

The contribution of the study is threefold: 1) Summarize and analyze the state-of-the-art of the literature in relation to KM, its application and relation to ASD, 2) Illustrate the KM strategies in the hierarchical layers of software development companies that adopt ASD, through a conceptual classification framework, and 3) Provide comprehension concerning how the companies implement KM strategies with practices that promote the KPs.

The rest of this paper is organized as follows: Section 2 introduces the background on ASD, KP, KM strategies and the conceptual classification framework. Section 3 describes the procedure for conducting this systematic literature review (SLR). Sections 4, 5, and 6 report the results, analysis, and discussions. In Section 7, we discuss threats to validity, and in Section 8 we provide conclusions and implications.

## 2. Background

In this section, we shortly present basic concepts of KM, explaining the types of knowledge, the knowledge processes, and the KM strategies. We provide a brief overview of ASD and a summary of the previous literature reviews in the area. We also compare them to our study, explaining the major differences.

### 2.1. Knowledge management

The challenge of allocating the knowledge asset logically is related to the dilemma of extracting and using knowledge that are distributed in many individuals' minds [20]. Although knowledge has its own theoretical definition, it is commonly used interchangeably with information. According to Nonaka [21], information is a collection of a particular arrangement represented by a flow of messages, whereas knowledge is the meaning created by the combination of information and individuals beliefs.

Knowledge is context dependent, dynamic and created through social interactions between individuals. Knowledge is divided in *tacit* – rooted in individual's mind and it is a result of values, beliefs, life experiences, emotions, procedures, actions and routines; and *explicit* – it is easy to transmit since it is already systematized in data, formula, manual, books, specifications, along with others [22] [23].

Even though knowledge is considered as the primary source of sustainable advantage for several companies [3], managing it remains challenging. Different fields have studied knowledge Management (KM) due to its relevance and interdisciplinary nature and proposed many concepts, theories, and applications. The term started to appear in academic publications in 1986 [24].

In this study, we adopt the **definition of knowledge management** (KM) as the effort to manage organizations' workforce through information and communication technologies or creation of a corporate culture that focuses on social processes that facilitates the sharing between individuals, aiming to reach a sustainable source of advantage [10][11].

KM strategy is the effort to formulate plans for actively manage knowledge [25]. The strategies can be *codification* and/or *personalization* strategies, which are rooted on the company's corporate strategy [16]. Codification involves the storage of knowledge into databases for further use (e.g., wiki-based tools [26]), while personalization focus on human interaction to communicate knowledge (e.g., communities of practice) [27].

The KM strategies promote the set of knowledge processes that represent individual's cognitive, social practices and the culture that shapes knowledge in organizations [9]. They are four KPs:

- *Knowledge creation* (KC) comprises the development of new ideas, concepts or knowledge replacement by the constant combination between tacit and explicit. A critical step to trigger this process is social interaction, which enables individuals to share and develop new knowledge [21]. In ASD, we could exemplify this process being triggered by pair programming, where developers elaborate create new knowledge to complex problems solution based on interaction;

- *Knowledge storage and retrieval* (KS) is related to the organizational memory, how it keeps the knowledge through documentation, databases, networks of individuals and so on. This memory is built with past experiences, events, and procedures that affect the organization's current activities. For example, inside agile teams, it is common utilize wiki-based tools to store additional documentation related to meetings and problems solution developed by the team. In SCRUM method, retrospective meetings could support KS;

- *Knowledge transfer/sharing* (KT) refers to transfer knowledge to areas needed within the organization. The knowledge disperses in different levels, between groups, individuals, across groups, and from the group to the organization. In this process, the main challenge is to know what knowledge the organization need to share, because usually, they do not know what they know. In ASD context, rotating individuals facilitate the transfer of knowledge between team members [28]. Wiki-based tools is another example of transferring knowledge about the software being developed from agile teams to different departments in a company [29];

- *Knowledge application* (KA) indicates the use of the knowledge as a competitive advantage through improvements in organization capability. Three mechanisms can integrate the knowledge: directives – the collection of rules and procedures; organizational routines – regards to coordination patterns and tasks development; and self-contained task – created especially for problem-solving, when there is no support from directives or organizational routines. In software development companies that adopt SCRUM framework, review sprint and retrospectives might be a good support for KA activities by effectively applying knowledge gained from previous sprints and projects.

A knowledge asset is a knowledge-based resource, for example, individual or team solutions developed for complex problems, related to feature development, based on individuals' previous knowledge and experience. Managing knowledge assets increases the return and also keeps the continuing advantages of new generated, transferred and applied knowledge [3][9]. In any organization, knowing the knowledge needs is one of the main steps for an effective KM, and also understand that filling repositories with knowledge is not the best approach to have favorable outcomes [30].

To process and apply knowledge is an intuitive activity, derived from mental mechanisms that are capable of connecting and merging conflicting knowledge [21][22]. Since knowledge is a large part of the resources used to build software products, it is crucial to know the cognitive structures process knowledge.

According to Robillard [31], there is a lack of understanding of how knowledge is processed in cognitive structures. Additionally, this process cannot be entirely automated. Different strategies are adopted aiming to codify the knowledge to automate managing it. On the other hand, several strategies focus on stimulating the cognitive structure of individuals through social interactions (personalization).

Depending on the organizational goals, a combination of these strategies may enhance one or more KPs. Considering that software development is a knowledge-intensive activity [32], that combines tacit and explicit knowledge during its lifecycle, it is valuable to understand how these constructs are interrelated to apply a good strategy combination.

## 2.2. *Agile software development*

Traditional software development processes struggle to offer the required flexibility and change responsiveness [33]. Promptly responding to changes is the prevalent idea of ASD [34] that had its principles disseminated in 2002. ASD principles arise as a contradiction to the traditional software development which is inspired by the project cycle and guided by tasks and documents [35]. Instead of extensive documentation to guide the product development, the ASD relies on the team capability of communicating, collaboration, and being creative [5].

Several ASD methods were suggested based on the principles established in 2001. Dybå and Dingsøyr [36] point out six primary agile methods: Crystal methodologies; Dynamic software development method (DSDM); Feature-driven development, Lean software development, Scrum, and Extreme programming (XP; XP2).

Successfully adopting ASD requires several adaptations, since it highly depends on the context that is applied. Contextual factors such as system size, customer demand for software, team expertise, and highly constrained quality attributes requires evaluation in agile adoptions [37]. A flexible organizational structure is also required for agile software development, where collaboration takes place.

Agile software development methods have been adopted in small collocated teams and also in large companies that have distributed teams [38][39][40]. Studies had already reported success on this adoption, although several challenges remain. Those challenges are related to team coordination, requirements definition, communication and knowledge sharing [41][42].

Traditional software development focuses mainly on explicit knowledge while in the agile context, the knowledge management activities are focused mainly on tacit knowledge [15]. Tacit knowledge is difficult to articulate since it is embodied in peoples' mind; it is a result of values, beliefs, and experiences [21]. In addition to this complexity, there are two notable obstacles for managing knowledge asset in software companies: time to transform the tacit knowledge into explicit and use it, and time pressure faced by project managers to deliver running code. However, benefits originated from KM could lead to a better competitive advantage when the knowledge has a crucial goal [12] [43] [44] .

### 2.3.    Previous systematic literature reviews

We identified five literature reviews that explore knowledge management and software development.

Bjørnson and Dingsøyr published a review [20] involving knowledge management in software engineering in 2008. They reported empirical studies on knowledge management in software engineering, including knowledge management concepts, major findings, and research methods. They observed that tacit knowledge is the focus of knowledge management activities in ASD, while explicit knowledge is more addressed in traditional development processes. They noticed that there were more lessons learned studies (or industrial case studies) than scientific contributions. Recommendations from the authors suggest that it is necessary to explore how to manage tacit knowledge.

In 2013, Camacho et al. published a systematic literature review [45] focusing on research on knowledge transfer in software engineering. They explore how to imply, how to measure, and how it occurs in ASD. Knowledge transfer is performed mainly through technology, using repositories to store information about the work that done and tools to increase collaboration between distributed sites. According to their findings, the conventional measures relates to the amount of innovation and number of new concepts successfully generated. It can also be measured with different rubrics, such as sales, customer satisfaction, number of customers, the value created by research and development activities, and return investment. They claim that the intensive communication facilitates knowledge transfer in the agile environment due to less documentation generation, the focus is on the interaction between people instead of processes and tools.

In 2014, Cabral et al. conducted a systematic literature review [46] about knowledge management in agile methodologies. They considered empirical and non-empirical studies in their analysis. They identified seven topics in the selected papers: project documentation, transfer and collaboration of tacit knowledge, adoption of knowledge management methodology, tools for knowledge management, human and social factor, use of communities of practice, and knowledge artifacts and experience knowledge. Regarding the software development cycle, the authors suggest that adopting knowledge management practices not only change the development cycle but also increase its costs. Most of the tools for knowledge management focus on communication between participants, which, for them, implies on the need for more studies where tools are more focused on the software development process.

In 2015, Ghobadi conducted a  systematic mapping study [25], to understand what drives knowledge sharing in software development teams [47]. She analyzed 49 papers and extracted 44 knowledge sharing drivers that were classified in four major categories (people-related, structure-related, task-related, and technology related), which had seven sub-categories (diversity-related drivers, capability-related drivers, team perception drivers, team organization drivers, team organization drivers, organizational practices drivers, task-related drivers, and technology drivers). The analysis suggests that the most cited drivers are the ones related to team perception, which refers to values, perception, and attitudes that drive the team to share; organizational practices, that refers to communication networks, organizational norms and practices; and technology-related, referring to templates, tools, and methodologies.

In 2017, Zahedi et al. published their review [48] about knowledge sharing challenges and practices in global software development. They grouped challenges and practices into six categories: management, team structure, work processes/practices, team cognition, social attributes, and tools. Most of the reported challenges regard the work processes, followed by social attributes and technology/techniques categories. The study reveals that one of the significant challenges faced by global software development is to deal with tacit knowledge and that there are costs involved in knowledge sharing, such as travel between sites and keep documentation up to date, that may not appear on the project planning. Among the reported practices, temporary collocation is the most popular knowledge practice used by the companies. Other practices are also reported, such as groupware tools to enable frequent communication and build social ties. In general, the majority of companies tend to influence the team's social potential to share knowledge.

### 2.3.1. Comparison between this study and the existing related literature reviews

From the five SLR's that we identified, three of them investigate knowledge sharing/transfer in different contexts of software development. One of them explores knowledge management in software engineering. The closest SLR to our work is the one reported by Cabral et al. [46] that aims to investigate the main topics of knowledge management in ASD. On a high-level, it seems similar. However, in detail, several differences can be perceived between this study and the other SLRs, regarding the following aspects:

- *Research questions*. The research questions elaborated in this study are different from all previous SLR's, (see Section 3.2). It may have an overlap of findings when comparing specific knowledge management processes such as knowledge sharing since the study published by Zahedi et al. [48] focus on challenges and practices of this KP. However, they consider this aspect in global software development. Our study explores all KPs in companies that adopt ASD.
- *Year*. Our search method, including automated search and snowballing procedure, was carried out on April 2018, while the closest to our study, reported by Cabral et al. [46], was performed in 2009, portraying a difference of eight years and 11 months. We consider this a reasonable timespan, considering that software development area has rapidly changed over the years, e.g., more companies are introducing agile methods, and several of them are also adopting it in global software development.
- *Search string*. The search string used in this study differs since did not adopt composite words for knowledge management and combine the words for agile software development differently. We also used different databases to run the search strings for the automated search; more is included, for example, Web of Science, that has better coverage in journals of high impact [49].

- *Search method*. Our study adopted two search methods, automated search and snowballing, while the other studies adopted mainly automated search. The automated search was used to find the start set for the snowballing procedure.
- *Inclusion and exclusion criteria*. We defined different inclusion/exclusion criteria, particularly the one related to empirical studies, through which we could aggregate different findings.
- *Quality assessment*. Since we were looking for empirical studies, we adopted the Rigor and Relevance model [50] to classify and evaluate the studies based on the rubrics for rigor and relevance to better judge the empirical level of the studies selected for analysis in our study.

Most importantly, to the best of our knowledge, this is the first study that focuses on KM processes in ASD and their association with KM strategies. Previous studies target different aspects of KM in ASD. Therefore, we believe this review contributes to a better understanding of how companies that adopt ASD manage the knowledge asset. These findings also highlight the role of an integrated approach to KM that consider not only the project level, but how these practices should be related to the company's overall goals.

## 3.    Research Methodology

We applied SLR to synthesize the empirical evidence regarding KM in ASD. In the subsections that follow we present the subsequent steps: protocol development, search process and study selection, quality assessment data extraction, and synthesis.

### 3.1.    Protocol development

We followed the guidelines proposed by Kitchenham [51] to structure the research question, search strategy, study selection, exclusion criteria, data extraction, and synthesis. Besides, we applied the Rigor and Industrial Relevance Model proposed by Ivarsson and Gorschek [50] to perform the quality assessment of the primary studies.

### 3.2.    Research question

We defined the research questions:

*RQ1: How do knowledge management strategies promote knowledge processes in ASD?*
*RQ2: To what degree have knowledge management strategies been validated in industrial settings?*
*RQ3: How have knowledge management practices been distributed in companies that adopted ASD?*

Motivation: Previous studies have focused in specific KPs, e.g., knowledge sharing [47][48], and related personalization strategies with knowledge sharing practices. However, the rest of the KPs (creation, storage, and application) remains uncovered regarding its practices and their relation to the KM strategy. As a consequence, our motivation relies on the need to gain comprehension about how software development companies adopting ASD implement KM strategies through practices that promote the KPs in the organizational layers, illustrated in the conceptual classification framework (see Section 5).

### 3.3. Search process and study selection

The search process followed two approaches. According to Kitchenhan et al., [51] and Wohlin [52], combining different approaches is a way of having the best possible literature coverage. We applied automated search and snowballing to select the primary studies.

The definition of the start set for the snowballing [52] procedure was carried out with the help of a database search. We searched for essential keywords and common terms, synonyms, and abbreviations in previous literature reviews [36][46][53]. Keywords related to agile software development were derived from Dybå and Dingsøyr [36], while keywords related to knowledge processes were not limited to composite words, for example, "knowledge creation" or "knowledge sharing". We decided to use only knowledge and learn variations (e.g., learning), due to the lack of information about the existence of studies that did not use combined words. By doing so, we reduced the precision of the search string, elevating the number of papers found (4951). Although, adopting this strategy we reduced the risk of missing relevant papers.

Table 1. Keywords for automated search

| Agile Software Development - (ASD) | Knowledge Management - (KM) |
|---|---|
| 1 agile AND software | 11 knowledge |
| 2 "extreme programming" | 12 learn* |
| 3 xp AND software | |
| 4 scrum AND software | |
| 5 crystal AND software AND (clear OR orange OR red OR blue) | |
| 6 dsdm AND software | |
| 7 "feature driven development" | |
| 8 fdd AND software | |
| 9 lean AND software | |
| 10 "dynamic system development method" | |

We combined all relevant keywords (refer to Table 1) using Boolean operators, and the search string result was:

(1 OR 2 OR 3 OR 4 OR 5 OR 6 OR 7 OR 8 OR 9 OR 10) AND (11 OR 12)

The search string execution and paper selection and screening were conducted in April 2018. We executed our search string in  Engineering Village that focus on engineering databases; Scopus, that is considered the most extensive citation and abstract databases; and ISI Web of Science, that has lower coverage than Scopus, but the journals covered has a higher impact [49].

*Automated search*

Fig. 1 presents the start set identification process (A1) on the databases (Scopus, Web of Science and Engineering Village) search was 4951. The paper's metadata retrieved from databases were stored using Zotero. A total of 1019 duplicates were removed, which left 3572 papers for further analysis.

At stage 2, the first author excluded papers not written in English and studies that did not undergo peer-review, resulting in 3222 candidates. On the third stage of the selection process, the first author excluded the papers that do not consider KM in ASD (exclusion criterion 3), resulting in 101 papers.

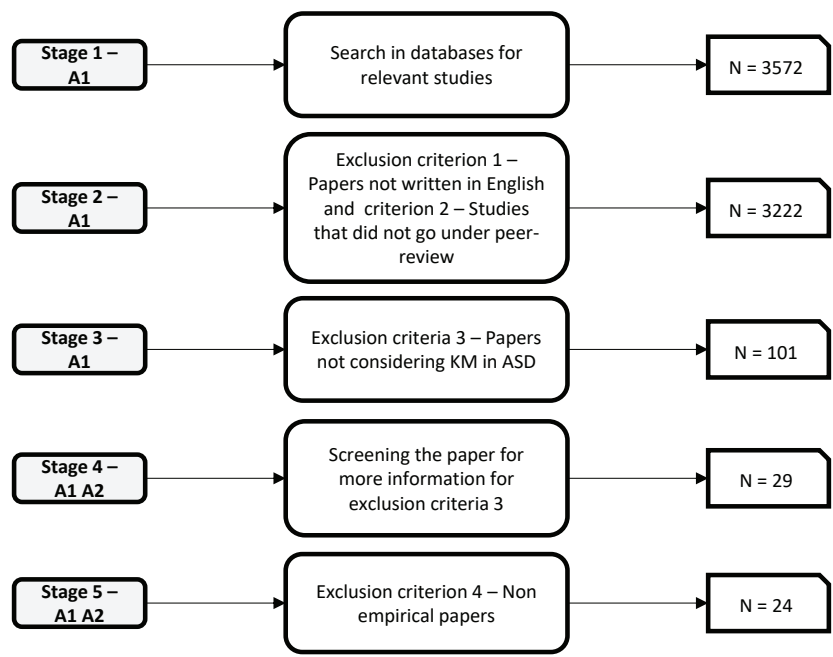

Fig. 1. Automated search for start set selection

Several papers did not present enough information that allowed us to exclude them by title or abstract. Because of that, the first author, together with the second author (Stage 4

– A1 A2) screened the remaining 101 papers and excluded 71 papers based on the third criterion.

At stage 5, the first and the second authors individually excluded papers based on criterion 4 – Non-empirical papers. The number of papers resulted from this stage was 24. In this stage, we calculated the Cohen's Kappa coefficient of agreement [54] to measure the degree of agreement between the researchers. The coefficient value was 0.87, which represents strong agreement [55].

*Snowballing iteration 1*

We performed backward and forward snowballing on the remaining 24 papers. Fig. 2 shows the process and the number of papers candidate for selection. We collected 2043 papers in the first snowballing iteration.

We first removed all papers that we already examined on the automated search. After that, by reading the titles and occasionally the abstracts, the first author selected 35 candidate papers. The first and the second authors individually screened the papers applying the third exclusion criteria, as mentioned earlier, to the 35 candidate papers. The Cohen's Kappa coefficient of agreement in this stage was 0.18, indicating a slight strength of agreement.

In several cases, it was not clear whether the context of the study was the agile environment, which explain the weak agreement. Due to the lack of information and to not make any assumptions, we decided to remove those papers. We run the selection once more, and in this round, the Kappa was 0.89, the strength of agreement almost perfect. Five papers were included and used in the first snowballing iteration.

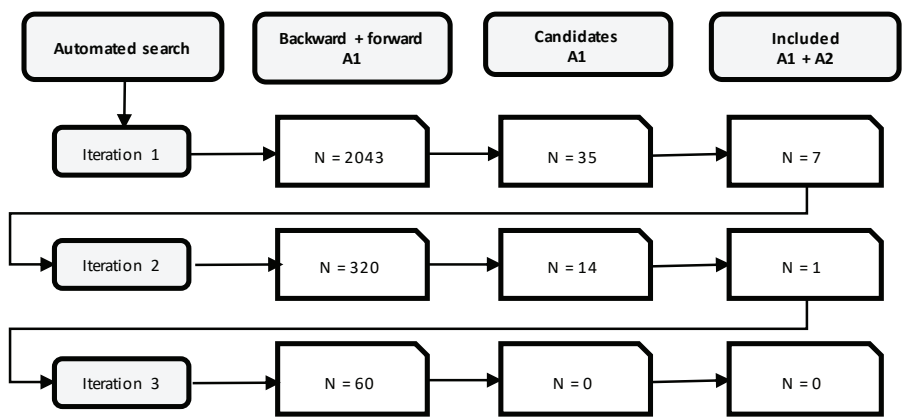

Fig. 2. Snowballing iterations

*Snowballing iteration 2*

In the second snowballing iteration (see Fig. 2), 320 papers were collected to analysis and 14 papers were considered as candidates. The first (A1) and the second (A2) authors individually evaluated the 14 papers using the exclusion criteria. For this step, the Kappa coefficient was 0.46, with moderate strength of agreement. One paper was included and used in the next iteration.

*Snowballing iteration 3*

For this iteration, we collected 60 papers, and none of them selected as candidates, concluding the snowballing part of the literature review. Overall, eight additional papers were included from the snowballing iterations, resulting in 32 papers to have their quality assessed before the data extraction begin.

### 3.4.    *Quality assessment*

We applied the quality assessment model based on their rigor and relevance scores proposed by Ivarsson and Gorschek [50]. Rigor aspects are related to three rubrics: context, study design, and validity. They were evaluated according to the respective scores, 0, 0.1, and 0.5. Relevance aspects have four rubrics, which are: context, research method, user/subject, and scale. The scores for these rubrics can be 0 or 1.

The paper's scoring was performed by the first and the second author. To validate this process both authors individually scored the papers and then they discussed the results of each one of them. We performed this discussion due to the possibility of having different perception about the model's rubrics and consequently the scoring process. After the discussion, the authors agreed on the final score for each paper (see Appendix B and Appendix C).

### 3.5.    *Data Extraction and synthesis*

The first author used a pre-defined spreadsheet for data extraction. Table 3 shows the data items collected.

We used quantitative analysis and descriptive statistic visualizations to explore and summarize the studies.

Table 2 Data collection of primary studies.

| Data collected | Research questions |
|---|---|
| Author, abstract, title, year of publication and type of article (conference, journal or workshop paper) | Demographic attributes for studies overview |
| Data collection method | Research design attributes |
| Rubrics related to rigor (context, study design description, and validity) | Study design attributes |
| Rubrics related to relevance (research design, subjects, context, and scale) | Study design attributes |
| Organizational layer of investigation | RQ1, RQ3 |
| KM practices | RQ1, RQ3 |
| Knowledge Processes | RQ1 |

To answer the research questions, we applied narrative synthesis [56]. We first classified the practices extracted from the primary studies in codification or personalization strategy, based on their definitions [27]. Second, we identified what KP these practices were enabling. The papers were clear on defining which KP they were aiming to investigate, e.g., knowledge sharing.

Finally, we examined in which organizational layer, described in the proposed framework (see Section 5) these practices are performed. Grouping the data into categories helped on exploring patterns and observe how KM practices are mapped in companies that adopted ASD.

## 4. Studies Overview and Descriptive Statistics

Fig. 3 depicts an overview of the 32 collected publications, including distribution through the years, publication venues, and type of method applied. We identified primary studies published in 2003 with a substantial increase in the number of publications after 2012 (see Fig. 3c).

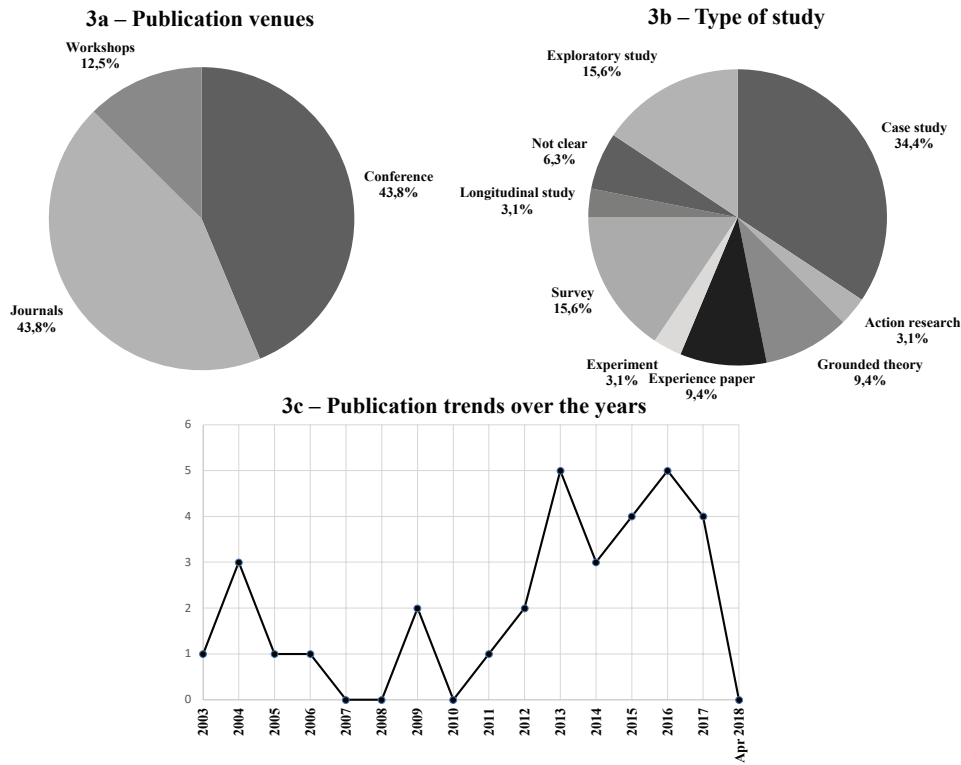

Fig. 3. Studies overview

We found 14 papers published in Journals, 14 in conferences and 4 in workshops, see Fig. 3a. The venues focus mainly on software engineering, computer science and information systems, except for two journals, *Knowledge and Process Management* and *VINE*, that focus on knowledge and information management.

We used the classification proposed by Robson and McCartan [57] to aggregate the studies according to their research methods (Fig. 3b). Case studies represent 34,4% (P1, P2, P3, P9, P10, P12, P18, P19, P26, P30, P31). Five exploratory studies were reported (P11, P20, P21, P22, P28), which had mainly interviews as data collection technique. Four studies report surveys (P13, P17, P23, P25, P32). Grounded theory (P5, P8, P24) and experience paper (P6, P15, P16) had the same amount of studies, representing each of them 9,4% of the studies. We also identified one action research (P4), one experiment (P7) and one longitudinal study (P14). Besides, we found two papers that did not indicate the applied research method (P27 and P29).

Overall, we observe that most papers present descriptive studies, which suggests a possible difficulty in conducting studies that require repeated observations along the time, e.g., longitudinal studies. Particularly in KM, real context experiments are difficult to conduct due to the complexity of setting up its environment, time for observation and changes in companies' routine. This possibly explains why we found only one experiment (P7), and it was conducted with students. Action research also demands time for progressive reflection with individuals within organizations, which aim to formulate solutions to companies' problems and to test them.

## 4.1. *Quality assessment*

To classify the papers based on the final score (refer to Section 3.4), we used the categories proposed by Munir et al. [58] to generate a grid of combinations. To score the papers, we applied the scale proposed by Ivarsson and Gorschek [50]. The four categories are:

- A – studies with high relevance and high rigor ($1,5 < rigor \leq 3 / 2 < relevance \leq 4$);
- B1 – studies with high relevance and low rigor ($0 \leq rigor \leq 1,5 / 2 < relevance \leq 4$);
- B2 – studies with low relevance and high rigor ($1,5 < rigor \leq 3 / 0 \leq relevance \leq 2$);
- C – studies with low relevance and low rigor ($0 \leq rigor \leq 1,5 / 0 \leq relevance \leq 2$).

The relevance aspect relates to the environment where the studies' results were obtained, more specifically, to the practical environment, industrial context, real applications and applied research method. Rigor concerns the scientific aspects of the reported research.

Fig. 4 depicts the distribution of the quality assessment scores. 16 papers were categorized in the B1 area (low rigor and high relevance). It suggests that relevant studies were conducted, but they should have been conducted with higher rigor to increase reliability and validity. One paper appeared in category B2 (P7), with high rigor and low relevance. In category C, three papers present low rigor and low relevance (P27, P4, 15). Category A is the second in the number of papers, with 12 (P8, P14, P19, P20, P21, P22, P23, P24, P25, P26, P28, P31), representing the papers with high rigor and high relevance.

Surprisingly, 59,3% of the studies were classified with low rigor regarding context, study design, and validity evaluation. When authors fail at describing research method and data analysis, reviewers and readers might misinterpret the research [50]. For practitioners, it is hard to analyze the studies' findings, since they might not be able to compare their context with the ones that describe poorly their context. Therefore, it is not possible to know if the practices adopted to manage knowledge are accurate and valid enough for practitioners' use. For researchers, the implications relate to how reproducible these studies are and how substantial is the evidence for future work.

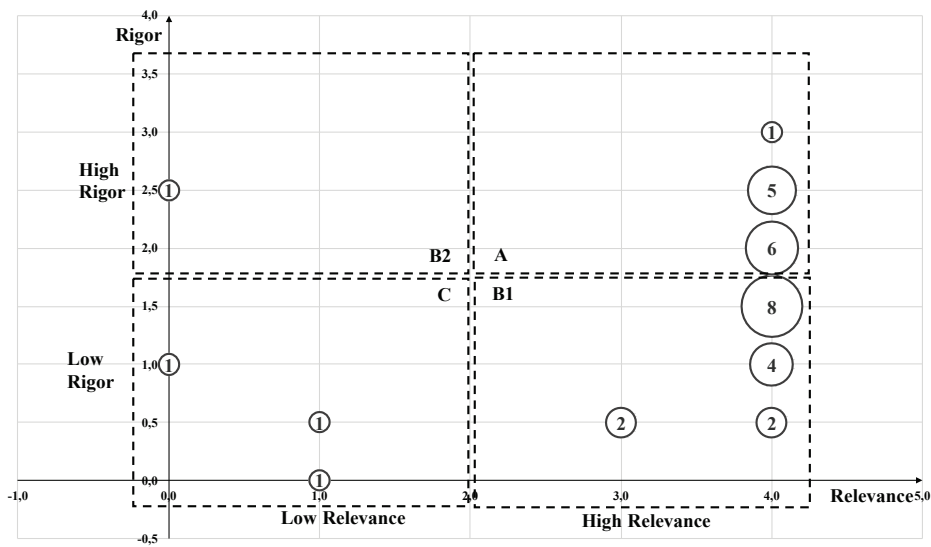

Fig. 4. Quality assessment

Regarding relevance, approximately 87,5% of the studies present high relevance. To this review, this result is significant since we are gathering findings from empirical studies in which industrial settings are more representative. On the other hand, 19 studies present low rigor, what weaken their reliability. They have been unsuccessful in reporting mainly the validity threats. The second rubric that present low scores was study design (see Appendix A), and third, the context description.

Knowledge is a complex and multifaceted theme; possible implications of low rigor on researching this theme could be the misinterpretation for both researchers and practitioners on what is considered knowledge by the companies. This understanding guides not only the research method to be used, but the companies' actions to manage individuals' knowledge. Increasing the studies rigor could contribute on filling this gap and provide stronger evidence that could support future works and more understanding for practitioners.

**5.      Results and Analysis**

To aggregate and analyze the results, we created The Knowledge Management Strategy in Agile Software Development (KMS-ASD) framework to explain how the different concepts are related, see Fig. 5.

*Framework conception*

Software development companies are often affected by changes to the external environment, market and customers' requirements. As a result, arrangements tend to change from hierarchical to flat, aiming more social interaction within and between organizational layers (strategic, tactical and operational); and changes in how an organization creates value through knowledge, e.g., factory model organization tends to have few knowledge-intensive functions, such as research and development departments [3]. Depending on the arrangements, the distance between the organizational layers may differ. With long distances, the translation of organization goals and requirements becomes a challenge, and may hinder its assimilation across the layers [59]. In the opposite direction, the communication from the lower to the other levels is also hampered.

The arrangements may also be different within the organizational layers, e.g., ASD companies have self-organized and greatly autonomous teams that rely on informal communication. Small to medium organizations, tend to have a flat structure of teams, which facilitates the implementation of KM strategies. Large organizations are often distributed, resulting in a more significant distance between the tactical and operational layers and greater team isolation.

Software development activities do not occur only in the bottom layer of a company, they relate to the other layers,  e.g., executives, product management or requirements engineering [60][61][62][63]. In the development process, the value created to the customers is supported by the organizational knowledge [64][65]; which is built by knowledge processes  within companies' routines.

The Knowledge Management Strategy in Agile Software Development (KMS-ASD) framework (refer to Fig. 5) envisions a high-level perspective of a software development company that has one or more software-intensive products and develops them through ASD processes. Vähäniitty and Rautiainen [60] propose a five-level conceptual framework (business, product & service, development portfolio, project, and iteration) that aims to link long-term product and business planning. We adapted their organizational levels by encapsulating business and product & service as a *Strategic layer*; naming development portfolio as a *Product portfolio layer*; and, product and iteration as a *Project layer*. This decision was made to simplify the framework and align it with the organizational structure proposed by Mahesh and Suresh [66], that has knowledge as central asset of an organization.

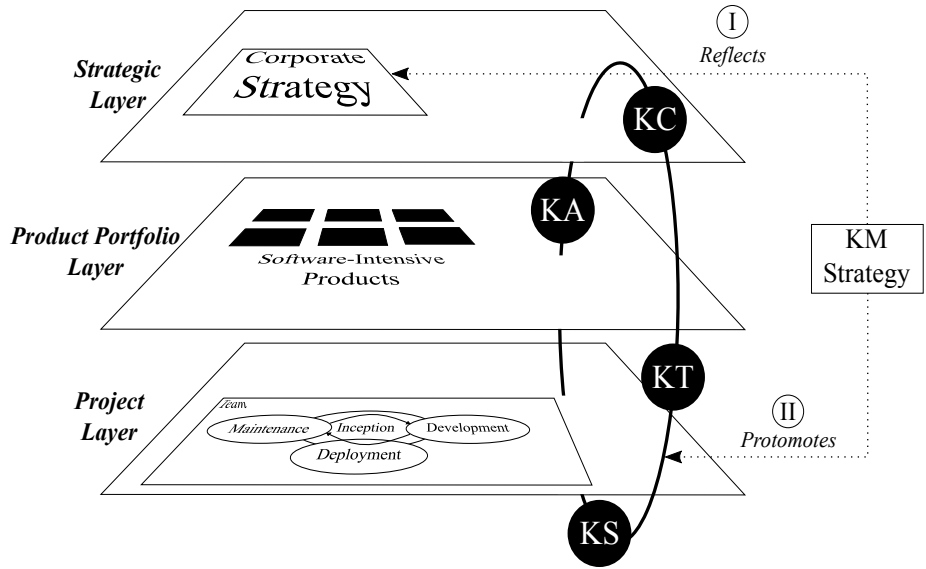

Fig. 5. KMS-ASD

In the KMS-ASD framework, KM strategies **reflect** (I) the corporate strategy, which defines how the company should compete for obtaining or maintaining competitive advantage [67] [68]. The KM strategies **promote** (II) at least one of the KPs [9][69] (KC - knowledge creation, KS – knowledge storage/retrieval, KT – knowledge transfer/sharing, and KA – knowledge application). The KPs take place within and between organizational layers. Codification and personalization strategies are usually combined; however, the distribution is not symmetrical, e.g., the mix could be, respectively, 80/20 for codification and personalization [27].

In the *Strategic layer*, the defined corporate strategy for obtaining competitive advantage is one input to define KM strategies [16]. The corporate strategy is discussed in terms of strategic positioning, the business model, current capabilities and required investments for new technologies and knowledge. The strategic level discussions also include how a company is going to achieve and sustain competitive advantage of market position.

The *Product portfolio layer* establishes the connection between strategic decisions and project levels. The strategic decision regarding the business, market positioning and the portfolio of product are taken here. The decisions about how to allocate the companies' resources into developing various products and what competences and knowledge are needed are also taken. In this level, personalization strategies that promote knowledge creation (KC) could be, for example, more aligned to the conception of a new product or product portfolio and their interaction with currently offered products or other services in an ecosystem. The product offering is developed concerning high level functionality and the product delivery strategy is agreed.

In the *Project layer*, the development iterations are planned according to the decisions made on the middle-level of the company [60]. Here, KM strategies are focused on practices that improve socialization between individuals inside and cross-teams. The less documentation approach to ASD foresee more interaction and, consequently, individuals disseminate more tacit knowledge.

The right balance of KM strategies could help knowledge storing and sharing from individuals throughout the company [12]. KM strategies may also focus on crossing layers. Launching new software-intensive products requires both technical and domain knowledge. The faster the new knowledge is spread and internalized, fewer delays may happen, and quality may be increased [12]. For companies that develop software connected or not to other products, this integration between KM strategies and the different levels could result in a successful competitive strategy.

In the following subsections, we list the KM practices reported in the primary studies regarding the KM strategy they implement (*RQ2*) and map them into the organizational layers. In each layer, we classify the practices regarding the type of KM strategy they implement, and which KP they promote (*RQ1*).

## 5.1. *Strategic layer*

Fig. 6 depicts the distribution of KM practices (*RQ3*) that implement both personalization and codification strategies in KMS-ASD.

We found no primary study that took into consideration the strategic layer regarding KM. A possible explanation for the absence of KM practices in this level could be that the performed practices are informal and not directly connected to the corporate strategy or goals. Practices in the strategic layer could assist in discovering new requirements to create new products or improve existing ones, oriented by the organizations' vision.

Knowledge is the leading resource for software development organizations and the integration among the different layers of the organization is essential. The reason for that is related to how the organizations create value. To Mahesh and Suresh [66], a flexible organization is recognized as an enabler of a free transit of knowledge by creating a social structure that systematically integrates knowledge from different layers.

Agile software development is suitable for this type of organizational structure due to the low level of formality in its development processes. Also, the encouragement of social interaction facilitates promoting the KPs, which might help ASD teams to rapidly adapt to changes not only in features, but in market demand as well. We believe that KM practices could be applied in ASD contexts and help to integrate knowledge from all layers and to support realizing the corporate strategy. For example, if a company aim to be leader in innovation, KM practices that promote knowledge creation might help on developing new products.

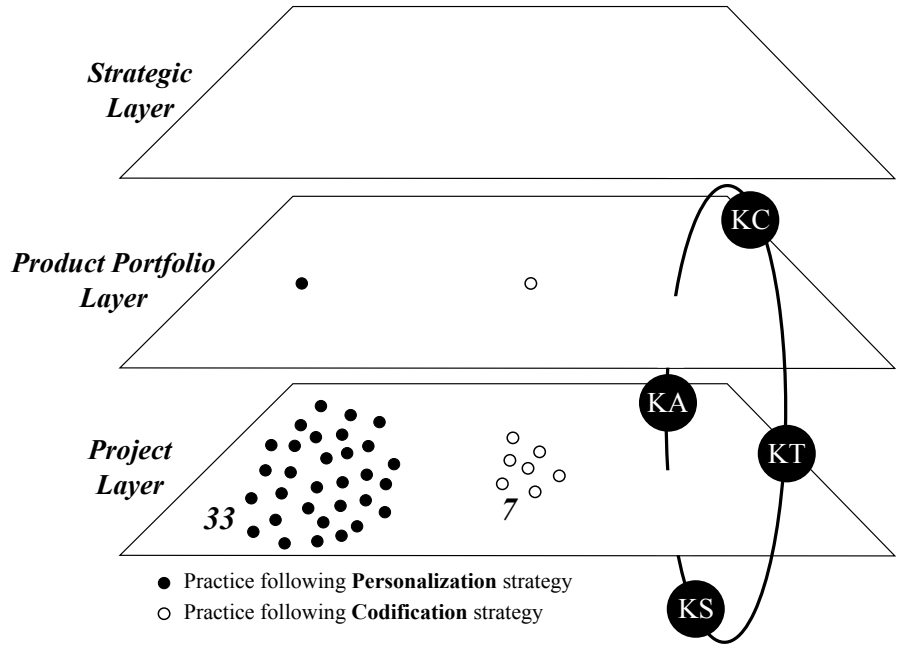

Fig. 6. Distribution of KM practices in organizational layers through KMS-ASD

However, based on our findings, it seems that agile teams tend to be isolated from the rest of the organizational layers, which means that KM practices could work well inside and across teams, but not integrated to the rest of the company. In a recent study [6], developers pointed that after agile adoption, they lost the "big picture" of the product because they are more focused on features do be developed. This finding could explain the lack of integration of KM practices to the other layers.

We suggest that future studies investigate how different organizational structures, e.g., flat and factory model, impact the KM practices integration to the different organizational layers of companies that adopt ASD. Moreover, use KMS-ASD framework to explain how strategies and practices are created and integrated to the strategic layer, mainly how they contribute to the organizations' competitive strategy.

The lack of integration may hamper the process of measuring the efficacy of the practices, due to impossibility of traceability of the practices in the project layer to the competitive strategy.

## 5.2.    Product portfolio layer

We identified two KM practices in the product portfolio layer: one implementing personalization strategy, and other implementing codification strategy.

As a personalization strategy, the *"marathon of innovation"* is a practice reported by Santos et al. (P24) that promotes knowledge creation. The company used its infrastructure to promote interaction among members from different teams and hierarchy levels to work together to stimulate innovation by developing new ideas. Individuals could, as a reward, dedicate time to maintain and grow the projects originated from their ideas.

The new ideas generation is a relevant topic and intrinsically related to KC. Santos et al. (P24) point out the relevance of the customer participation and the role of companies in enabling customer participation. Customers play a key role in ASD since they provide information about requirements and contribute to socializing their needs and experience, which may trigger the KC process to result in new solutions.

Regarding the codification strategy, *wiki-based tools are used to transfer knowledge between different departments*. This practice was identified in one study (P6), and allows the connection between different business units, keeping knowledge and information updated. The authors also point out that the tool was intensively used by the developers to store the product documentation and knowledge considered valuable by them.

Both practices focus innovation and knowledge sharing between product portfolio and project layer. To establish a connection beyond project barriers, patterns of communication and knowledge transfer are required. Strode et al. [70] identified in their study that companies use "boundary spanning artefacts" to enable this communication.

This type of artefact could be helpful on propagating knowledge through organizational layers. However, one main challenge remains unaddressed regarding this aspect, which is to know what knowledge the organization needs to share on what level and across the levels.

### 5.3. *Project layer*

We identified 40 KM practices that aim attention at the project layer. Regarding personalization strategies, the primary studies report practices that promote KC, KT and KA processes. Codification strategy gathers seven practices, they promote KS, KT and KA processes.

#### 5.3.1. *Personalization strategies*

The KM practices that implement personalization strategies are divided into the following knowledge processes: Seven in knowledge creation; 25 in knowledge transfer/sharing; and one in knowledge application.

**Knowledge creation process**. Knowledge is created through KM practices that stimulate socialization of tacit knowledge. Three of the reviewed studies (P5, P22, P1) report KM practices for KC. All three studies have high relevance, but studies P1 and P5 have low scores for study design, discussion about validity threats, and context description. It hampers the comparison to other studies' results and the understanding about on other similar context KC practices could be suitable; for instance, the comparison between P5 and P22 regarding the context where communities of practices are applied.

Bahli and Zeid (P1) argue that the software development process adopted intervene on KC. After conducting an empirical study in industry, the authors concluded that the *adoption of extreme programming* facilitated KC.

*Formal and informal learning practices* reported in one primary study are applied to stimulate knowledge creation during project execution (P5). According to Dorairaj et al. (P5), these practices aim to improve team members' skills by upgrading their technical and management knowledge. *Communities of practices* (P5, P22), *pair programming* (P22), and *innovation boards* (P22) can be used to stimulate social interaction between the team members during software development. The *customer interaction* is a KM practice reported as a relevant source of KC (P5, P22). To Razzak and Ahmed (P22), daily customer involvement throughout the development cycles allow team members to discuss issues and create knowledge through continuous feedback.

**Knowledge transfer process**. 11 primary studies report the application of different communication channels that stimulate social interaction (P2, P5, P12, P13, P16, P18, P20, P22, P24, P26, P29). In distributed teams, this interaction is even more challenging. To minimize the distance and time zone differences, *information and communication tools*, such as video conferencing, chat rooms and telephones are adopted (P29, P20, P26). Wendling et al. (P29) argue that technology is essential to connect distributed teams. However, it does not replace face-to-face interaction. *Moving members between sites for a short time* complements this practice. Boden and Avram (P2) found out that giving the opportunity for *team members to work together for long periods* in a distributed context, facilitates KT between members from different sites.

Collocating teams tend to reduce the demand for information and communication tools. The focus changes to promote a friendly environment that encourages KT. *Communities of practices, workshops, consultancy, frequent meetings*, and *informal gatherings* are practices used to share knowledge among individuals (P2, P5, P12, P13, P16, P20, P22, P24, P26, P31). To Karlsen et al. (P12), the interaction during the meetings promotes discussions about different subjects, tasks, solutions, and estimations. This practice allows team members to share knowledge to build solutions together and discuss their progress and goals. In a case study at Ericsson, Šmite et al. (P26) found that communities of practices contribute to develop and increase the knowledge network by the frequent communication between teams. Santos et al. (P25) report one practice called *brainwriting*, a practice similar to brainstorming. In this case, the ideas were written in papers. According to the authors, a brainwriting session focus on reusing knowledge for problem-solving and on innovating by creating new concepts.

The physical working structure is a concern reported in one primary study. Santos et al. (P24) explain that the physical workspace affects KT effectiveness. Walls that separate offices act as barriers for KT. For them, *open workspaces* that integrate people and furniture that facilitate working together (e.g., tables for pair programming) provide closeness between members.

Companies adapt the software development process and team configuration to achieve team cohesion and better structure to share tacit knowledge (P5, P8, P12, P18). Ramesh et al. (P18) point that the time established for the development cycles depends on the complexity of the functionalities that the team is working. Because of that, the development *cycles are short, but not time-boxed*. In the studied companies, the *length of the sprints was* *changed*, aiming to achieve more interaction for KT. The authors also point that *keeping the same team during the whole project* is a practice that benefits team cohesion and KT as well.

Besides the adopted development process, the team configuration also undergoes adjustments. Two studies (P5, P12) reported the adoption of *cross-teams* configuration. To Karlsen et al. (P12), tacit knowledge transfer may be facilitated when the team is composed of individuals that assume different roles and have diverse backgrounds. Besides, an increase of motivation is also observed as result of cross-teams' configuration.

Practices that increase trust between individuals and long-term collaboration have a positive effect on KT in agile teams (P2, P8, P16). To Moe et al. (P16), *conceive and discuss product goals collectively*, mainly when the teams' members are gathered, induced the members to understand that they need to trust each other and share knowledge to achieve the goals. This practice is also pointed by Ghobadi and Mathiassen (P8) as valuable for knowledge sharing.

The primary studies report practices that aim to leverage individuals´ knowledge through interaction, resulting in technical knowledge sharing. Studies (P5, P22, P24) report that companies promote *formal* or *informal technical discussion sessions* to discuss technical solutions. In Santos et al. (P24), one company promote *challenge activities* support learning through interaction, aiming to share and improve programming skills.

*Pair programming* is reported as a beneficial practice in both collocated and distributed teams. The informal conversation during the development offers a good opportunity for the team members to share their knowledge (P7). In an empirical study, Ghobadi and Mathiassen (P8) found that when there is a lack of understanding about the business domain, *to engage experienced and motivated people* could help to share knowledge. Another practice that aims to level up knowledge between individual is the *rotation of members*, which is applied in collocated or distributed teams (P5, P12, P20, P22, P24). The rotation occurs by changing the individuals' roles from one team to another or rotating team member between sites when the company has distributed development. In Santos et al. (P24), most of the studied companies adopt rotation of members when they need to structure new teams.

The relation of customer with the teams is explored in two studies (P5, P8). In Ghobadi and Mathiassen (P8), *leveraging a good relationship with the client* and *organizing training sessions* with the client are practices that provide opportunities to transfer knowledge in both ways. Dorairaj et al. (P5) suggest that having an *on-site customer* during the whole development helps on providing not only feedback but also collaborating throughout the iterations.

The concept of high-quality knowledge sharing is introduced by Ghobadi et al. (P7), and is related to how useful is the knowledge transferred to the software development activities. In an experiment, they explore how *coopetitive reward structures* influence the high-quality knowledge. The authors claim that on one hand, cooperation leads to better performance; on the other, competition generates more analytical thinking. By combining them, they found that there is a direct influence on the level of high-quality knowledge.

Six of the studies (P8, P20, P22, P24, P25, P26) received high rigor and relevance scores. In this case, the findings regarding the practices utilized for KT may be considered more consistent, providing better support for replication and practical relevance.

Seven studies (P2, P5, P12, P13, P16, P18, and P29) received high relevance but low rigor scores. Validity threats discussion and study design have the lowest scores. The implication of those scores for KT practices is the impossibility of establishing an appropriate comparison between the contexts. This fact may be more critical for distributed environments, since the contexts may differ significantly. Effectively describing the context where KM takes place may help in comparison and understanding about the frequency of social interaction, goals for transferring knowledge, people involved, type of knowledge, and transfer success rate.

Understand how each company perceive the concept of knowledge is crucial for the reproducibility of the KT practices, which is not successfully reported in the primary studies. The lack of rigor could imply in misinterpretation for researchers and subjects about the concepts and practices goals. Particularly, P7 is evaluated with high rigor, but with low relevance, mainly because of using students as subjects in the experiment. Besides, the applications used for the experiment are unrealistic, do not represent real industry context.

**Knowledge application process**. This process expresses how the knowledge is transformed into competitive advantage of a company. We found only one study that reports one practice that assists KA in the project layer. Dorairaj et al. (P5) report that the *sprint review* practice in Scrum helps to identify areas that need improvement for the next sprint and helps to create favorable circumstances for new ideas generation.

We could not find practices enabling KS. One possible explanation is that practices to store and retrieval knowledge are highly dependent on explicit knowledge, which is not the objective of the personalization strategy.

### 5.3.2.  *Codification strategies*

We found seven KM practices that follow codification strategies: three of them are related to KS, three to KT and one to KA. According to Alavi and Leidner [9], KM practices that implement codification strategies aim to store knowledge through documentation or databases. The goals are to share explicit knowledge, to build up experience,  and to contribute with new experience e knowledge.

In **knowledge transfer** process, we found two studies (P4, P8) that report documentation of project experiences. Dingsøyr and Hanssen (P4) propose a *lightweight approach to post-mortem reviews*. The teams discuss and document good and bad experiences from the projects and involve stakeholder in the review process. According to Ghobadi and Mathiassen (P8), *documented experience* supports planning future projects by observing inadequate planning performed in the past.

*Information and communication technologies* are widely adopted by companies studied in the primary studies (P5, P12, P29, P13, P22). Companies keep track of technical details of the software, project documents and meetings to share with the teams. Wiki-based tools are the most used, together with the tools for knowledge transfer, such as JIRA, ScrumWorks, Confluence, Hudson, Yammer, Github, electronic boards and Redmine. One study reports *visual prototyping* as a practice to share knowledge (P20); however, we could not find any detail about how this practice is conducted.

We found three studies in the **knowledge storage process** area (P3, P5, P18, P32). Ramesh et al. (P18) report the case of a company that developed a *database* that their teams use to store different types of information and knowledge, to report issues and get feedback on solutions. The database is also reported by Gervigny and Nagowah (P32) to store lessons learned and good work practices.

*KM systems* are used to store content from technical presentations, concepts and technical expertise for further consideration by team members. Software details are stored in the code itself (P5).

Chau and Maurer (P3) developed an *online tool to store knowledge* based on the wiki technology. It is associated with a database capable of storing structured and unstructured information. Besides, the tool aims to support the experience factory approach through structured knowledge and experiences of learning, which is similar to communities of practice.

We found one codification practice in the **knowledge application process** area. Dorairaj et al. (P5) studied a company that documents knowledge from projects in a wiki-based tool, aiming to *reuse* or adapts in future projects in similar context. The teams document knowledge related to issues and solutions, adding detailed explanations.

We were not able to find practices that promote the KC process. One possible explanation is the nature of the KC process, that is about externalizing tacit knowledge to combine it with other explicit knowledge. To do that, tacit knowledge requires being externalized to be stored.

Two papers received low rigor and relevance scores (P3, P4). In P4, despite the use of real industry projects, they use students subjects; this scenario does not represent a real industry environment. In P3, the practice of using a tool is undermined because of the lack rigor on validating the tool in a real industry context. Therefore, these practices should be considered carefully by practitioners and researchers for two main reasons: lack of context description and industry representation.

## 6.    Discussion and Research Gaps

In the primary studies, we observe that companies combine both personalization and codification strategies. 81% of the KM practices implement personalization (promote social interaction between team members) while 19% of the practices implement codification. This means that the practices of codification strategies act as support for the personalization strategies.

Regarding the KPs promoted by the KM strategies, we observe a tendency on the **KT** practices to focus on tasks and problem solutions by enhancing face-to-face or digital communication. However,  the primary studies do not explicitly explain what knowledge is transferred. Using informal communication creates a risk of losing knowledge or not transferring it. The result is that knowledge gets localized inside a few individual's minds instead of being propagated.

We found only three empirical studies that focus on **KC**. This indicates that more in-depth studies are required to understand why ASD needs to promote KC and what knowledge they aim to create. KC is not a trivial process to investigate, and probably the reason of the low number of primary studies found is because it requires substantial time from researchers to observe how the knowledge was created, shared and resulted in a new product, process improvement, solutions, among other outcomes.

Curiously, the **KA** process is the least investigated by the reported studies. The organizational competitiveness may be affected by the ability to effectively apply knowledge [17]. Future studies focusing on ASD could approach where knowledge is applied and how it has become a solution or a product.

In **KS**, we notice that software companies still struggle to know what knowledge they should store. More important than having a tool, is to know what to store regarding information or knowledge. Additionally, researchers should explore how one can adapt storage and retrieval activities in ASD, more specifically, when these activities occur in the project and what are the guidelines the teams should follow to perform them.

To summarize, it is surprising that the identified papers do not discuss KM practices efficacy. Are these practices effective? A previous study conducted in 2008 also presented this concern [15]; however, our review reveals the same lack of studies regarding the efficacy of KM practices.

We recommend to apply empirical and rigorous methods for investigating KM in ASD, such as case studies, followed by in-depth interviews and observational studies since the beginning of a project till it ends, or even longitudinal studies. By doing this, it is possible to trace the knowledge throughout the development process, its externalization, combination, and internalization inside/across the teams/company. Another viable research method is the ethnographic study, also suggested by Bjørnson and Dingsøyr [15].

## 7.    Threats to Validity

External validity remains a significant concern since we studied the intersection of ASD and KM. We consider that our findings are generalizable to companies that adopt ASD principles through the different methodologies in software development. However, a risk remains that there could exist studies in ASD or KM literature that could be considered relevant but do not explicitly consider both areas. Due to the systematic nature of this study, we consider this risk to be low.

Concerning internal validity, we address the five characteristics: study selection, subjectivity on quality assessment, misclassification of primary studies, no exclusion of papers with low rigor/relevance, and database selection.

In the **study selection** phase, disagreements may have led to the exclusion of relevant papers. To mitigate this risk, the two first authors independently applied the exclusion/inclusion criteria and measured the related agreement through the Cohen's Kappa coefficient [54].

The second characteristic is a degree of **subjectivity on quality assessment**. Although the model [50] used to evaluate the quality of the primary studies presents rubrics with clear description, understanding them and scoring each paper could introduce misjudgment on quality evaluation. In this case, the first and the second author performed this process together, discussing the model rubrics and how the primary studies satisfied them.

Referring to the **misclassification of primary studies**, the lack of proper description about KP and KM strategies described by the authors in some of the investigated primary studies could have jeopardized the reliability of our results. To mitigate that, we defined all theoretical concepts and terminology for both constructs, ASD and KM, and classified the papers during the synthesis base on these definitions.

Regarding the decision of **not excluding primary studies** with low rigor/relevance, it could have weakened the evidence of our findings. However, excluding these papers, we could miss relevant or rigorous findings, mainly because all of them report empirical studies. Since they already underwent peer-review, we believe that we keep a balance between volume and validity of our findings.

To avoid the risk of missing papers that are not indexed by the **selected databases**, we complemented the database search with the snowballing procedure. This approach helps reconsidering papers that were excluded by mistake on previous steps, since several of them emerge again during the procedure.

## 8. Conclusions and Implications

Comprehending KM concepts becomes more important as companies start to consider knowledge as one of the main resources for building products. Successful KM strategies implementation involves knowing what goals the company aim to achieve, and its relation to the corporate strategy. Thereafter, the KM strategy promote the KPs through practices in the different layers of a company.

By following a systematic procedure, we conclude:

- KM strategies in ASD promote mainly KT process with practices that stimulates social interaction to informally share tacit knowledge;

- The primary studies reported practices in product portfolio and project layers. Most of the practices (95%) are placed in the project layer and the studies do not report about the connection of KM strategies and practices between the strategic layers and product portfolio or project layer;

- 81% of the reported KM practices implement personalization strategy, while 19% of the KM practices implement codification strategy (acting as a support for personalization).

- Personalization strategies promote KC by stimulating close interaction between team members and customers. Personalization strategies also promote KT by stimulating social interaction inside and between teams, modifying the agile method adopted to accommodate knowledge transfer practices, and customer participation; and promote KA process by using review sprint approach.

- Codification strategies support storing of the project documentation, learning materials, and meeting notes. KT process is supported by several documents, i.e., post-mortem reviews and prototyping. The codification strategies promote KS using databases and information systems, i.e., wiki; and promote KA by using the knowledge stored in wikis.

The results of this review have implications for both researchers and practitioners. For researchers, we identified several potential research gaps in Section 6. We summarize our indications for future research into three main implications: 1) to investigate how KM strategies are planned in ASD; 2) to explore how organizational structures (layers and size) impact KM strategies implementations in ASD context; and 3) How to define measures for KM practices' efficacy in ASD.

For researchers, the results could contribute to a better comprehension of how the strategies are related to the processes through KMS-ASD framework, and on which gaps future research should focus. For practitioners, this study offers several insights for KM practices in ASD context, depending on what KP they mainly want to promote. The combination of KM strategies shows that personalization strategies are predominant in ASD. The framework could guide practitioners in planning KM activities that support the corporate strategy at all organizational layers, considering what knowledge should be transferred from and to the different layers.

### Acknowledgements

This work is supported by the IKNOWDM project (20150033) from the Knowledge Foundation in Sweden.

**Appendix A. Selected papers**

**Appendix B. Rigor evaluation**

| Paper ID | Reference | Context (C) | Study Design (S) | Validation (V) | C+S+V |
|----------|-----------|-------------|------------------|----------------|-------|
| P1 | [18] | 0.5 | 0.5 | 0.5 | 1.5 |
| P2 | [28] | 0.5 | 0.0 | 0.0 | 0.5 |
| P3 | [71] | 0.5 | 0.5 | 0.5 | 1.5 |
| P4 | [72] | 0.0 | 0.0 | 0.0 | 0.0 |
| P5 | [43] | 1.0 | 0.5 | 0.0 | 1.5 |
| P6 | [29] | 0.5 | 0.0 | 0.0 | 0.5 |
| P7 | [73] | 1.0 | 1.0 | 0.5 | 2.5 |
| P8 | [13] | 1.0 | 1.0 | 0.0 | 2.0 |
| P9 | [74] | 0.5 | 1.0 | 0.0 | 1.5 |
| P10 | [75] | 0.5 | 0.5 | 0.0 | 1.0 |
| P11 | [76] | 0.5 | 0.5 | 0.0 | 1.0 |
| P12 | [77] | 0.5 | 0.5 | 0.0 | 1.0 |
| P13 | [19] | 0.5 | 0.5 | 0.5 | 1.5 |
| P14 | [78] | 1.0 | 1.0 | 0.5 | 2.5 |
| P15 | [79] | 0.5 | 0.0 | 0.0 | 0.5 |
| P16 | [80] | 0.5 | 0.0 | 0.0 | 0.5 |
| P17 | [81] | 0.0 | 0.5 | 0.0 | 0.5 |
| P18 | [82] | 1.0 | 0.5 | 0.0 | 1.5 |
| P19 | [83] | 1.0 | 1.0 | 0.5 | 2.5 |
| P20 | [84] | 1.0 | 1.0 | 0.0 | 2.0 |
| P21 | [85] | 1.0 | 1.0 | 0.5 | 2.5 |
| P22 | [86] | 1.0 | 1.0 | 0.5 | 2.5 |
| P23 | [87] | 0.5 | 1.0 | 0.5 | 2.0 |
| P24 | [17] | 1.0 | 1.0 | 0.5 | 2.5 |
| P25 | [88] | 0.5 | 1.0 | 0.5 | 2.0 |
| **P26** | [89] | 1.0 | 1.0 | 1.0 | 3.0 |
| **P27** | [90] | 0.5 | 0.5 | 0.0 | 1.0 |
| **P28** | [91] | 1.0 | 1.0 | 0.0 | 2.0 |
| **P29** | [92] | 1.0 | 0.5 | 0.0 | 1.5 |
| **P30** | [93] | 0.5 | 0.5 | 0.0 | 1.0 |
| **P31** | [94] | 1 | 1 | 0 | 2 |
| **P32** | [95] | 0.5 | 0.5 | 0.5 | 1.5 |

## Appendix C. Relevance evaluation

| Paper ID | Reference | User/ Subject (U) | Context (C) | Scale (S) | Research Method (RM) | U+C+S+RM |
|---|---|---|---|---|---|---|
| **P1** | [18] | 1.0 | 1.0 | 1.0 | 1.0 | 4.0 |
| **P2** | [28] | 1.0 | 1.0 | 1.0 | 1.0 | 4.0 |
| **P3** | [71] | 0.0 | 0.0 | 0.0 | 1.0 | 1.0 |
| **P4** | [72] | 0.0 | 0.0 | 0.0 | 1.0 | 1.0 |
| **P5** | [43] | 1.0 | 1.0 | 1.0 | 1.0 | 4.0 |
| **P6** | [29] | 1.0 | 1.0 | 1.0 | 0.0 | 3.0 |
| **P7** | [73] | 0.0 | 0.0 | 0.0 | 0.0 | 0.0 |
| **P8** | [13] | 1.0 | 1.0 | 1.0 | 1.0 | 4.0 |
| **P9** | [74] | 1.0 | 1.0 | 1.0 | 1.0 | 4.0 |
| **P10** | [75] | 1.0 | 1.0 | 1.0 | 1.0 | 4.0 |
| **P11** | [76] | 1.0 | 1.0 | 1.0 | 1.0 | 4.0 |
| **P12** | [77] | 1.0 | 1.0 | 1.0 | 1.0 | 4.0 |
| **P13** | [19] | 1.0 | 1.0 | 1.0 | 1.0 | 4.0 |
| **P14** | [78] | 1.0 | 1.0 | 1.0 | 1.0 | 4.0 |
| **P15** | [79] | 1.0 | 0.0 | 0.0 | 0.0 | 1.0 |
| **P16** | [80] | 1.0 | 1.0 | 1.0 | 0.0 | 3.0 |
| **P17** | [81] | 1.0 | 1.0 | 1.0 | 1.0 | 4.0 |
| **P18** | [82] | 1.0 | 1.0 | 1.0 | 1.0 | 4.0 |
| **P19** | [83] | 1.0 | 1.0 | 1.0 | 1.0 | 4.0 |
| **P20** | [84] | 1.0 | 1.0 | 1.0 | 1.0 | 4.0 |
| **P21** | [85] | 1.0 | 1.0 | 1.0 | 1.0 | 4.0 |
| **P22** | [86] | 1.0 | 1.0 | 1.0 | 1.0 | 4.0 |
| **P23** | [87] | 1.0 | 1.0 | 1.0 | 1.0 | 4.0 |
| **P24** | [17] | 1.0 | 1.0 | 1.0 | 1.0 | 4.0 |
| **P25** | [88] | 1.0 | 1.0 | 1.0 | 1.0 | 4.0 |
| **P26** | [89] | 1.0 | 1.0 | 1.0 | 1.0 | 4.0 |
| **P27** | [90] | 0.0 | 0.0 | 0.0 | 0.0 | 0.0 |
| **P28** | [91] | 1.0 | 1.0 | 1.0 | 1.0 | 4.0 |
| **P29** | [92] | 1.0 | 1.0 | 1.0 | 1.0 | 4.0 |
| **P30** | [93] | 1.0 | 1.0 | 1.0 | 1.0 | 4.0 |
| **P31** | [94] | 1 | 1 | 1 | 1 | 4.0 |
| **P32** | [95] | 1 | 1 | 1 | 1 | 4.0 |